\documentclass[aps,prl,groupedaddress]{revtex4}
\usepackage{amsmath}
\usepackage{epsfig,epsf}
\usepackage{graphicx}
\usepackage{verbatim}
\usepackage{epstopdf}

\newcommand{\be}{\begin{equation}}
\newcommand{\ee}{\end{equation}}
\begin{document}

%\hoffset-1cm

% Yale printer values
\voffset1.5cm
%
%\documentclass[prb,12pt]{revtex4}

%\usepackage{amsmath}    % need for subequations
%\usepackage{graphicx}   % need for figures
%\usepackage{verbatim}   % useful for program listings
%\usepackage{color}      % use if color is used in text
%\usepackage{subfigure}  % use for side-by-side figures
%\usepackage{hyperref}   % use for hypertext links, including those to external documents and URLs
%\newcommand{\be}{\begin{equation}}
%\newcommand{\ee}{\end{equation}}
%\begin{comment}
%\pagestyle{empty} % use if page numbers not wanted
%\end{comment}

%\usepackage{epsfig,epsf}
%\usepackage{graphicx}
%\usepackage{grffile}

%\begin{document}

%\hoffset-1cm

% Yale printer values
\voffset1.5cm

%\draft{BI-TP 2005/19}
%\preprint{BI-TP 2005/19, CERN-PH-TH-2005/99}
\title{The Curious Case of an Effective Theory.}
\author{ Ibrahim Burak Ilhan and Alex Kovner}

\affiliation{
Physics Department, University of Connecticut, 2152 Hillside
Road, Storrs, CT 06269-3046, USA}
%\date{\today}

\begin{abstract}
We describe an effective theory of a scalar field, motivated by some features expected in the low energy theory of  guodynamics in 3+1 dimensions. The theory describes two propagating massless particles in a certain limit, which we identify with the Abelian QED limit, and has classical string solutions in the general case. The string solutions  are somewhat unusual as they are multiply degenerate due to spontaneous breaking of diffeomorphism invariance. Nevertheless all solutions yield identical electric field and have the same string tension.
\end{abstract}
\maketitle
% above is the preamble

\section{Introduction}
The aim of this note is to describe a field theoretical model, motivated by an attempt to extend the understanding of certain aspects of 2+1 dimensional gauge theories to 3+1 dimensions. 

In 2+1 dimensions one has a very simple and straightforward relation between confinement and spontaneous breaking of a discrete magnetic symmetry\cite{thooft},\cite{dualreview}. Additionally, in 2+1 dimensions, nonabelian gauge theories, exhibiting the phenomenon of confinement, are related to abelian theories on the effective field theory level, by a simple simmetry breaking deformation. This deformation breaks the continuous $U(1)$ magnetic symmetry of an abelian theory down to a discrete group $Z_N$ for $SU(N)$ gauge theories with adjoint matter. The mere fact of the presence of this deformation, coupled with the spontaneous breaking of the residual discreet group leads with certainty to a confining long distance behavior\cite{dualreview},\cite{chris}. 

In this note we discuss a 3+1 dimensional model that displays similar features. The model is designed to implement certain aspects of abelian-nonabelian transition, similar to 2+1 dimensional case. Although it clearly cannot be taken literally as the effective theory of QCD, curiously enough it does have some similarity with Fadeev-Niemi model, that has been proposed as an effective theory of glueballs from a completely different perspective\cite{niemi}.

Before discussing the model itself, we will recall briefly the story of 2+1 dimensional gauge theories. As a prototypical Abelian gauge theory consider scalar QED. It posesses a continuous $U_\mu(1)$ global symmetry generated by the total magnetic flux through the plane of the system, $\Phi=\int d^2x B(x)$. The order parameter  
for this symmetry is one complex field $V$, which creates pointlike magnetic vortices. In the Coulomb phase $\langle V\rangle=v \ne 0$ and $U_\mu(1)$ is spontaneously broken. The low energy dynamics is qualitatively described by the effective "`dual"' Lagrangian 
\begin{equation}\label{l}
L=-\partial^\mu V\partial_\mu V^*-\lambda (V^*V-\frac{e^2}{8\pi})^2
\end{equation}
The Goldstone boson of the $U_\mu(1)$ symmetry breaking is identified with the massless photon, while the electric charge in the dual formulation is the topological charge of the field $V$
\begin{equation}
J_\mu=\frac{1}{e}\epsilon_{\mu\nu\lambda}\partial_\nu V^*\partial_\lambda V
\end{equation}
A charged state of QED in the effective description appears as a hedgehog - like soliton of $V$: $V(x)=ve^{i\theta(x)}$, with $\theta=\tan^{-1} y/x$.

This effective formulation is also a good basis for description of confinement in nonabelian theories. In particular the effective theory of a weakly interacting $SU(N)$ model is essentially the same as eq.(\ref{l}), except with a potential which breaks the magnetic $U_\mu(1)$ symmetry down to $Z_N$
\begin{equation}\label{lN}
L=-\partial^\mu V\partial_\mu V^*-\lambda (V^*V-\frac{e^2}{8\pi})^2+\mu(V^N+V^{*N})
\end{equation}
The perturbation reduces the infinite degeneracy of vacua of the Abelian theory to a finite number of degenerate vacuum states connected by the $Z_N$ transformations. As a result, a charged state does not have a rotational symmetry anymore, but the winding is concentrated within a quasi one dimensional  "`flux tube"'\cite{dualreview}. 

This is a very simple picture, and a very appealing one inasmuch as it identifies charged objects with topological defects which inherently have long range interactions due to their topological nature. It also identified photons with Goldstone bosons, providing a natural symmetry based explanation for masslessness of the photon.

It is natural to ask, whether in 3+1 dimensions one can have a similar description, which encompasses  the massless nature of photons in QED as well as topological mechanism of confinement in nonabelian theories. The situation here of course, is much more complicated. First of all, in 3+1 dimensions photons are vector particles and so it is not clear at all whether they can be understood as Goldstone bosons. Even if such a case can be made for photons, it is not easy to identify the relevant conserved current that breaks spontaneously. It is clear that the current has to be related to the dual field strength $\tilde F_{\mu\nu}$ consistent with the fact that photons have spin one\cite{photongoldstone}. The dual field strength, however has no local order parameter, and thus is an object of a very different kind than ordinary vector currents, which we are used to deal with. Another complication is, that classical effective description assumes weakly interacting theory, while QCD is of course strongly interacting. 

All these are difficult questions, to which we do not attempt to provide answers here. Instead we will be content to construct a model that encompasses the following basic features: 

1. The model should describe dynamics of scalar fields, and contain no fundamental gauge fields.

2. The model should have the limit (putative "Abelian regime") in which it has two massless degrees of freedom, which are identified as Goldstone bosons. These massless Goldstone bosons in our model are intended to play the role of photons.
 
  3. In the Abelian regime the model must provide for existence of classical topological solitons, which play the role of electrically charged particles. We require the topological charge that is carried by these solitons to reflect the mapping of the spatial infinity onto the manifold of vacua, and thus be given by $\Pi_2(M)$. The energy of these solitons has to be finite in the infrared. The energy density of a soliton solution should decrease as $1/r^4$ far from the soliton core. This is nontrivial in 3+1 dimensions, since our model has no gauge fields, while scalar fields that contribute to $\Pi_2$ have to be long range.  
  
  4. Soliton must become confined in the "Nonabelian regime", when a symmetry breaking perturbation is added. This same perturbation must eliminate massless Goldstones by expicitly breaking the (previously) spontaneously broken symmetry group down to a discreet subgroup. Confinement should be accompanied by formation of string between the solitons.
 
 In this note we present a model which has all the above features and discuss its properties, which are somewhat unusual. In particular, the requirement of the finiteness of the energy of a topological soliton in the Abelian regime is very restrictive. It leads to rather unusual properties of the confining strings in the Nonabelian regime such as existance of an infinite number of zero modes. This degeneracy can be lifted, however that requires the addition of another perturbation which is not clearly related with breaking of symmetries of the theory.
 
We stress, that the model we discuss does not provide perfect emulation of many properties of gauge theories. In particular, even in the Abelian regime, it contains classical solutions with magnetic charge density, and thus the effective dual field strength tensor is not conserved. Related to that, although we are able to construct solutions of equations of motion that behave as single photons, the model has no solutions that correspond to a two photon state with arbitrary polarization vectors. Nevertheless, we think that the model has many similarities with gauge theories, and thus is sufficiently interesting to be considered perhaps as a simplified prototype for future work.

\section{The Abelian Model.}
\subsection{The Field Space and the Lagrangian.}
As explained in the introduction, we wish to construct a model of scalar fields which contains two massless degrees of freedom and solitons of finite energy. The simplest option that we adopt is a theory with two scalar degrees of freedom endowed with $SU(2)$ symmetry. Spontaneous breaking of this symmetry must lead to two massless modes.  Thus we choose as the configuration space the $O(3)$ nonlinear $\sigma$-model.
\begin{equation}
\phi^a, \ a=1,2,3; \ \ \ \ \phi^2=1
\end{equation}
The moduli space allows for the requisit topology $\Pi_2(S^2)$. The topological charge associated with it, is identified with the electric charge of QED
\begin{equation}\label{topch}
Q=\frac{e}{4\pi^2}\int d^3x\epsilon_{abc}\epsilon_{ijk}\partial^i\phi^a\partial_j\phi^b\partial_k\phi^c
\end{equation}
As the first task, we have to contend with the following potential problem. In a theory with the standard kinetic term, the energy of a state with nonvanishing topological charge diverges in the infrared. A typical topologically nontrivial field configuration is a rotationally symmetric hedgehog
\begin{equation}\label{hedge}
\phi_h^a(x)=\frac{r^a}{|r|}f(|r|); \ \ \ \ \ \ \ \ \ f(|r|)\rightarrow_{r\rightarrow\infty}1
\end{equation}
The standard two derivative kinetic energy diverges quadratically on such a configuration. In order to make the energy of the soliton finite, we need to introduce a kinetic term with more than two derivatives. 

In fact, there exists a unique four derivative term which is a natural choice for a kinetic term for our model. The identification of the electric charge with the topological charge eq.(\ref{topch}) also naturally leads to the identification of the electric current as
\begin{equation}
J^\mu=\frac{e}{4\pi^2}\epsilon_{abc}\epsilon^{\mu\nu\lambda\sigma}\partial_\nu\phi^a\partial_\lambda\phi^b\partial_\sigma\phi^c
\end{equation}
and therefore electromagnetic field tensor as
\begin{equation}\label{field}
F^{\mu\nu}=\epsilon^{abc}\epsilon^{\mu\nu\lambda\sigma}\phi^a\partial_\lambda\phi^b\partial_\sigma\phi^c
\end{equation}
Since our goal is to construct a model that resembles QED as close as possible, the natural choice for the kinetic term is the square of the field strength tensor, which is just the well known Skyrme term. 

Hence we consider the model of a triplet of scalar fields defined by the following Lagrangian:
\begin{equation}\label{ourl}
L=\frac{1}{16e^2}F^{\mu\nu}F_{\mu\nu}+\lambda(\phi^2-1)^2
\end{equation}

We note, that the sign of the $F^2$ term in the Lagrangian is opposite to that in QED. In the framework of eq.(\ref{ourl}) the sign is determined so that the Hamiltonian is positive, rather than negative definite. This feature is common to models related by duality. For example the same is true in  the 2+1 dimensional models described in the introduction, where the kinetic term in the Lagrangian of the effective theory when written in terms of the field strength tensor has the opposite sign to that of Electrodynamics. The reason for this inversion, is that while in QED the electric field is proportional to the time derivative of the basic field (in this case $A_\mu$), in the effective dual description it is the magnetic field that contains time derivative of the vertex field $V$. Thus in order for the Hamiltonian of the two models in terms of $E$ and $B$ to be the same, the Lagrangians have to have opposite sign. This inversion of sign also takes place in our model, and is the natural consequence of the relation between the field strength tensor and the basic scalar degrees of freedom eq.(\ref{field}).

In the strongly coupled limit $\lambda\rightarrow \infty$, the isovector $\phi$ has unit length, and the field strength is trivially conserved
\begin{equation}
\partial_\nu F^{\mu\nu}=0
\end{equation}
This limit therefore corresponds to QED without charges. In this limit the energy of the soliton eq.(\ref{hedge}) diverges linearly in the ultraviolet. At finite coupling $\lambda$ the variation of the radial component of the field $\phi^a$ softens the UV behavior, and the soliton energy is UV finite. It is also IR finite thanks to our choice of the four derivative action. 
In fact on the hedgehog configuration eq.(\ref{hedge}) the " electric field" decreases as $E_i(x)\propto \frac{r^1}{|r|^3}$, and the energy density away from the soliton core decreases as $1/r^4$, just like the Coulomb energy of a static electric charge in the electrodynamics.

\subsection{The Equations of Motion}
We now derive the equations of motion for the model. For conveniene we  define in the strong coupling limit
\begin{equation}
\phi_3 = z, \ \ \ \ \ \ \psi = \phi_1 + i \phi_2 = \sqrt{1-z^2} e^{i \chi}
\end{equation}
With this parametrization one has
\be F^{\mu \nu} = \epsilon^{\mu \nu \alpha \beta} \epsilon_{abc} \phi^a \partial_ \alpha \phi^b \partial_\beta \phi^c = - 2 \epsilon^{\mu\nu\alpha\beta}\partial_\alpha z \partial_\beta \chi\ee
The Lagrangian can be written as 
\be L=\frac{1}{4e^2}(\partial_\mu z\partial_\nu \chi-\partial_\mu\chi\partial_\nu z)^2\ee
The equations of motion read
\begin{eqnarray}
&&\partial^\mu\Big[\frac{1}{e^2}\partial^\nu\chi\left(\partial_\mu z\partial_\nu \chi-\partial_\nu z\partial_\mu \chi\right)\Big]=0\nonumber\\
&& \partial^\mu\Big[\frac{1}{e^2}\partial^\nu z\left(\partial_\mu z\partial_\nu \chi-\partial_\nu z\partial_\mu \chi\right)\Big]=0
\end{eqnarray}
This can be combined into
\begin{equation}\label{eqmo}
\frac{1}{e^2}\partial_\nu G(z,\chi)\partial_\mu\left(\partial_\mu z\partial_\nu \chi-\partial_\nu z\partial_\mu \chi\right)=  \frac{1}{e^2}\partial_\nu\Big[ G(z,\chi)\partial_\mu\left(\partial_\mu z\partial_\nu \chi-\partial_\nu z\partial_\mu \chi\right)\Big]=0
\end{equation}
where $G(z,\chi)$ is an arbitrary function of two variables. These equations have a form of conservation equations for currents defined as
\begin{equation}\label{current}
J_\nu^G=G(z,\chi)\partial^\mu\left(\partial_\mu z\partial_\nu \chi-\partial_\nu z\partial_\mu \chi\right)
\end{equation}
\subsection{The Symmetries and the Correspondence to Electrodynamics.}
The conserved currents of eq.(\ref{current}) can indeed be identified with conserved Noether currents. An unexpected consequence  of the choice of the Skyrme term as the kinetic term in the Lagrangian, is that the global symmetry group of the model is much larger than the $SO(3)$ group we have started with.  

To see this note, that the field strength as defined in eq.(\ref{field}) is related to an infinitesimal area on a configuration space. Let us be more precise here. A given field configuration $\phi^a(x)$ defines a map from space-time to a sphere $S^2$. Consider a given component  the field strength tensor, say $F_{12}$ at some point $x$. To calculate it in terms of $\phi$ we consider three infinitesimally close points $A\equiv x^\mu$, $B\equiv x^\mu+\delta^{\mu 1}a$ and $C\equiv x^\mu+\delta^{\mu 2}a$. These three points in space-time, map into three infinitesimally close points on the sphere $\phi^a(A),\ \phi^a(B),\ \phi^a(C)$. The field strength $F_{12}$ is proportional (up to the factor $a^{-2}$ to the area of the inifnitesimal triangle on $S^2$ defined by these three points. 
Since the action of our toy model depends only on $F^{\mu\nu}$, it is clear that any reparametrization of the sphere which preserves area is an invariance of our action. 

Thus the $SO(3)$ global symmetry we started with, is a {\it small} subgroup of the area preserving diffeomorfisms of $S^2$, which we denote $Sdiff(2)$\cite{sdiff}. This is the group of canonical transformations of a classical mechanics of one degree of freedom.
 The infinitesimal symmetry transformation in terms of $z$ and $\chi$ is
\begin{equation}\label{sdiff}
z\rightarrow z+\frac{\partial G}{\partial \chi}; \ \ \chi\rightarrow \chi-\frac{\partial G}{\partial z}
\end{equation}
with arbitrary $G(z,\chi)$.
The appropriate Noether currents are precisely those of eq.(\ref{current}) and the equations of motion are indeed equivalent to conservation equations of these currents.

It is amusing to note that this symmetry is similar to the world sheet diffeomorphism invariance of the Nambu-Gotto string. Indeed, if one thinks of the fields $z$ and $\chi$ as the world sheet string coordinates, the world sheet diffeomorphism invariance is precisely eq.(\ref{sdiff})\cite{misha}. Although our setup looks very different from a string theory, there may be more to this analogy than meets the eye, as the basic "`order parameters"' of the magnetic symmetry in $QED_4$ are indeed magnetic vortex strings\cite{photongoldstone}. The  $S^2$ topology of the world sheet then implies closed string loops. We will not develop this analogy any further here, and instead will return to the field theoretical approach.

The enhanced symmetry  means that the moduli space is much larger  than $S^2$ as would be naively the case for symmetry breaking pattern $SO(3)\rightarrow SO(2)$. Any configuration $\phi^a(x)$ that maps the configuration space into an arbitrary {\bf one dimensional} curve on $S^2$ has vanishing action and is thus a point on the moduli space. The moduli space is therefore the union of maps $\phi^a(x)$ that map $R^4$ to  $L$, where $L$ is an arbitrary point or a one dimensional curve on $S^2$. 

Nevertheless, even though the moduli space is not a simple sphere, the topological charge $Q$ is quantized for any smooth classical configuration of fields $\phi(x)$. A twist in the tale is that there are many more degenerate soliton configurations than just the rotationally invariant hedgehog of eq.(\ref{hedge}). Any $Sdiff(2)$ transformation coresponding to an arbitrary regular function $G$ of eq.(\ref{sdiff}) applied to the configuration eq.(\ref{hedge}) generates a soliton configuration $\phi^{aG}_h(x)$ which is degenerate in energy with $\phi^a_h(x)$. Note, that although these are different field configurations, they all correspond to the same electric field $E_i=\epsilon_{ijk}\epsilon^{abc}\phi^a\partial_j\phi^b\partial_k\phi^c$, since the electric (as well as magnetic) field is invariant under the action of $Sdiff(2)$ transformations.
\subsection{Plane waves -  photon states.}
Returning to the Lagrangian eq.(\ref{ourl}), the natural question to ask is how much of a relation does it have with electrodynamics. With the identification eq.(\ref{field}), we know that the field strength $F^{\mu\nu}$ satisfies half of Maxwell's equations. The equations of motion eq.(\ref{eqmo}) are quite reminsicent of the other half of Maxwell's equations. They can be rewritten in terms of $F^{\mu\nu}$ as
\be\label{max}[\partial_\nu G(z,\chi)]\partial_\mu \tilde F^{\mu\nu}=0\ee
Thus, any configuration of the fields $z,\ \chi$ that satisfies $\partial_\mu \tilde F^{\mu\nu}=0$, also satisfies the equations of motion of our model. The converse is not true: there are solutions of the equations of motion eq.(\ref{eqmo}) which do not satisfy the equations of motion of electrodynamics. We give an example of such a solution in the Appendix.

The model eq.(\ref{ourl}) is therefore not equivalent to electrodynamics. Nevertheless, it is interesting to ask whether the spectrum of solutions of eq.(\ref{ourl}) contains basic excitations of QED, in particular the photons. This is a slight abuse of language, since we are dealing with a classical theory. We will nevertheless refer to plane wave configurations of $F^{\mu\nu}$ with lightlike momenum as photons.

Our aim in this section is to show that the free wave excitations are indeed solutions of equations eq.(\ref{eqmo}).  To this end consider the configuration
\be \label{photons} \chi(x)=A\epsilon^\mu x_\mu; \ \ \ \ \ \ z(x)=\sin k^\mu x_\mu\ee
where the vector $\epsilon^\mu$ is normalized as $\epsilon^\mu\epsilon_\mu=-1$.
On this configuration 
\be \label{photons1}\tilde F^{\mu\nu}=A(\epsilon^\mu k^\nu-\epsilon^\nu k^\mu)\cos k\cdot x\ee
Thus
\be \partial_\mu \tilde F^{\mu\nu}=-A\Big[(\epsilon\cdot k)k^\nu-k^2\epsilon^\nu\Big]\sin k\cdot x\ee
This vanishes, provided the momentum vector is lightlike and the polarization vector $\epsilon$ is perpendicular to $k$:
\be \label{photons2} k^2=0; \ \ \epsilon\cdot k=0\ee
For a given lightlike momentum $k_\mu$ this equation has three independent solutions for $\epsilon^\mu$. One of them, however is proportional to $k_\mu$ itself. With this polarization vector, the field strength tensor vanishes. Thus there are two independent polarization vectors $\epsilon^\mu_\lambda,\ \ \lambda=1,2$ that correspond to plane wave solutions for $F^{\mu\nu}$. Just like in QED, it is convenient to choose the polarization vectors so that their zeroth component vanishes $\epsilon_\lambda^\mu=(0,\epsilon^i_\lambda)$.  
The constant $A$ is the overall amplitude of the electromagnetic wave, whose square is porportional to the number of photons with a given momentum and a given polarization in the wave. 

The arbitrariness in the choice of the polarization vectors is precisely the same as in the choice of polarization vectors in Electrodynamics
\be \epsilon^\mu\rightarrow \epsilon^\mu+ak^\mu\ee
Note that this shift of polarization vector is affected by the transformation
\be \chi\rightarrow \chi+a\arcsin z\ee
which is one of the $Sdiff(2)$ transformations eq.(\ref{sdiff}). In fact the two field configurations eq.(\ref{photons}) can be transformed by any element of $Sdiff(2)$ without changing $F^{\mu\nu}$.

The solution eqs.(\ref{photons},\ref{photons1},\ref{photons2}) describes a state in all respects equivalent to the freely propagating photon, and we will refer to it as such. The setup here is dual to the usual free QED. Normally one introduces the vector potential $A_\mu$ via $F^{\mu\nu}=\partial_\mu A_\nu-\partial_\nu A_\mu$. This relation potentiates the homogeneous Maxwell's equation $\partial_\mu\tilde F^{\mu\nu}=0$. However, in the free chargeless QED entirely analogously one can potentiate the other Maxwell equation by introducing the dual vector potential via $\tilde F^{\mu\nu}=\partial_\mu \tilde A_\nu-\partial_\nu \tilde A_\mu$. The dynamics of the dual vector potential $\tilde A_\mu$ is identical to that of $A_\mu$, and it can be expanded in exactly the same polarization basis as $A_\mu$. In this formulation QED pocesses a dual gauge symmetry $\tilde A_\mu\rightarrow \tilde A_\mu+\partial_\mu\lambda(x)$.

To make the correspondence between our model and the Electrodynamics more transparent, we can define a dual vector potential
\be\label{dualv} \tilde A_\mu=z\partial_\mu \chi\ee
Under the $Sdiff(2)$ transformation  eq.(\ref{sdiff}) it transforms as
\be \label{transf}\tilde A_\mu\rightarrow \tilde A_\mu +\partial_\mu[G-z\frac{\partial G}{\partial z}]\ee
which has a form reminiscent of the dual gauge transformation in Electrodynamics with the gauge function $\lambda(x)=G-z\frac{\partial G}{\partial z}$.

The analogy of eq.(\ref{dualv}) with the dual vector potential of QED is suggestive, but one has to be aware that this is only an analogy rather than equivalence. First, the tranformation eq.(\ref{transf}) is not a gauge transformation, but rather the action of  a global symmetry transformation of the Lagrangian on $\tilde A_\mu$ of eq.(\ref{dualv}). More importnantly, the vector field defined in eq.(\ref{dualv}) in terms of two scalar fields is not the most general form of a vector field, even allowing for a possible gauge ambiguity. For that reason the variation of the Lagrangian with respect to such a constrained vector potential does not lead to the homogeneous Maxwell's equation directly, but instead to eq.(\ref{max}), which allows additional solutions.

Finally we note, that the solution eq.(\ref{photons}) gives a nice and simple interpretation for the properties of the photon states in terms of the effective theory. The photon momentum is the momentum associated with the variation of the third component of the isovector $\phi^a$, while the direction of the photon polarization vector is determined by the spatial variation  of the phase $\chi$.

Although a plane wave $\tilde F_{\mu\nu}$ solves the equations of motion, the equations eq.(\ref{max}) are not linear in the basic field variables, and thus an arbitrary superposition of two such solutions, itself is not a solution. We may try to construct a two photon state by slightly extending the ansatz eq.(\ref{photons}).
\be\label{photons3}  \chi = \lambda_\mu x_\mu; \ \ \ z= a \sin{k^\mu x_\mu}+ b\sin{p^\mu x_\mu}\ee
with $k^\mu$ and $p^\mu$ - both lightlike vectors, $\lambda^\mu k_\mu=\lambda^\mu p_\mu=0$ and $\lambda^\mu \lambda_\mu=-1$.
The latter two conditions can be satisfied by taking
\be\lambda^\mu=\alpha\Big[\epsilon^\mu- \frac{\epsilon \cdot k}{k \cdot p} p_\mu - \frac{\epsilon \cdot p}{k \cdot p} k_\mu \Big]\label{lambda}\ee
with an arbitrary vector $\epsilon^\mu$ and an appropriate normalization constant $\alpha$.

 The dual field strength tensor can be written as:
\be \tilde{F}_{\mu \nu} = a (k_{\mu}\epsilon^k_\nu-k_\nu\epsilon^k_\mu)
 \cos{k\cdot x}  +b (p_{\mu}\epsilon^p_\nu-p_\nu\epsilon^p_\mu)
 \cos{p\cdot x} \ee
with the polarization vectors
\be \epsilon^k_\mu=\lambda_\mu- \frac{\lambda_0}{k_0}k_\mu; \ \ \ \epsilon^p_\mu=\lambda_\mu- \frac{\lambda_0}{p_0}p_\mu;\ee
This is a bona fide two photon state. Unfortunately by varying $\lambda_\mu$ at fixed $p$ and $k$ one cannot obtain two most general polarization vectors for photons with momenta $k$ and $p$. This is obvious since both polarization vectors $\epsilon^k$ and $\epsilon^p$ have equal component in the direction perpenducular to the plane spanned by $p^i; \ k^i$. Thus we are lacking one degree of freedom to be able to construct a two photon state with arbitrary polarizations of both photons. In the Appendix we show that this is problem is not restricted to our ansatz eq.(\ref{photons3}), but is a genuine limitation of our Lagrangian.

\section{The Nonabelian perturbation and the string solution.}
In analogy with 2+1 dimensions we now perturb the theory with the simplest perturbation which breaks the $O(3)$ global symmetry. This perturbation should eliminate the vacuum degeneracy associated with the spontaneous symetry breaking. We find it convenient to choose a potential that fixes the vacuum expectation value of the field $z$ to be equal to unity. We thus consider the Lagrangian
\be L=\frac{1}{16e^2}F^{\mu\nu}F_{\mu\nu} - \frac{2}{e^2}\Lambda^2 (z-1)^2\ee
The perturbation breaks not only the $SO(3)$ symmetry, but also generic $Sdiff(2)$ transformations.  Nevertheless, the subgroup generated by 
\be\label{subgroup}
\chi\rightarrow\chi -\frac {dG(z)}{dz}
\ee
remains unbroken. We keep this in mind throughout the discussion of this section.

The equations of motion now are 
\begin{eqnarray}
&&\partial^\mu\Big[\frac{1}{e^2}\partial^\nu\chi\left(\partial_\mu z\partial_\nu \chi-\partial_\nu z\partial_\mu \chi\right)\Big]=\frac{4}{e^2}\Lambda^2(z-1)\nonumber\\
&& \partial^\mu\Big[\frac{1}{e^2}\partial^\nu z\left(\partial_\mu z\partial_\nu \chi-\partial_\nu z\partial_\mu \chi\right)\Big]=0
\end{eqnarray}

With this perturbation there are no finite energy solutions with nonvanishing topological charge $Q$. Instead, we expect to find translationally invariant stringlike solution. In the presense of a soliton and antisoliton such strings will connect the two and will provide for a linear confining potential. To find such a solution consider a static field configuration translationally invariant in the third direction. For such a configuration  the only non-vanishing components of $F_{\mu\nu}$ are:

\be F^{0 3 } = 2\epsilon^{ij}\partial_i z \partial_j \chi\ee

We look for a  solution invariant under rotations in the plain perpendicular to the string
\begin{equation}
\chi(x)=\theta(x); \ \ \ \ \ \ \ \ z(x)=z(r)
\ee
where $r$ and $\theta$ are the polar coordinates in the $x_1,x_2$ plain.
Such a configuration has a unit winding in the $x_1,x_2$ plain which is precisely what one expects from the string connecting the soliton and antisoliton. The soliton, resides at some very large negative value of $x_3$. far to the left of the soliton the field configuration must be vacuum $\phi^1=\phi^2=0; \ z=1$. Thus the topological charge calculated on a surface enclosing such a soliton is equal to the two dimensional topological charge - the winding of the phase $\chi$ on any plain pierced by the string. An identical argument applies for the antisoliton, which resides at large positive value of $x_3$. Thus indeed our ansatz is appropriate for the string connecting a soliton and an antisoliton residing far apart. For our ansatz the equation of motion for the field $\chi$ is trivially satisfied. The equation for $z$  becomes

%\be 32 \frac{1}{r}[\partial_\theta \chi\partial_r[\frac{1}{r}(\partial_r z \partial_\theta \chi - \partial_\theta z \partial_r \chi)] - \partial_r \chi \partial_\theta [\frac{1}{r}(\partial_r z \partial_\theta \chi - \partial_\theta z \partial_r \chi)]] \label{zettam}\ee

%\section{Equations with $\Lambda$-term}

%Then, the equations of motion are:

%\be -\frac{1}{4}\partial_s \frac{\partial F^2}{\partial (\partial_s z)} + 2 \Lambda (z-1) \ee

%Let's try to find a solution such that $\chi = \theta$, and $z = z(r)$. In this case, the $\chi$ equations are trivial, and the $z$ equation is:

%\be - 2 \frac{1}{r} \partial_r [\frac{1}{r}(\partial_r z)] + 2 \Lambda^2 (z-1) = 0 \ee
\be 4z''=4\Lambda^2(z-1)\ee
where $z'\equiv \frac{dz}{dr^2}$

%\be 4 \partial^2_{r^2} z - \Lambda^2 (z-1) = 0\ee

For the solution to be well defined in the middle of the string and approach vacuum far away from it, $z$ must have the asymptotic behavior:

\be z(0)=-1, \mbox{   } z(\infty)=1\ee

The solution is easy to find

\be z(r^2) = 1 - 2 \exp\{-\Lambda r^2\} \label{zetsifir}\ee

The solution has some intuitively expected properties: it has a finite width, outside which the fields approach vacuum, while inside the string the field values are different from the vacuum and thus it carries a finite energy density. The string tension can be calculated exactly
\be \sigma=8\pi\frac{\Lambda}{e^2}\ee
Nevertheless, the solution is rather peculiar, since it does not approach the vacuum exponentially, but rather as a Gaussian. the string therefore has a very sharply defined width, outside of which the vacuum is reached very quickly. In a theory with a finite mass gap and a finite number of massive excitations such behavior is impossible. 
This strange behavior can be traced back to our noncanonical kinetic term. Indeed, for simple dimensional reasons, the kinetic energy for a rotaionally invariant configuration is a second derivative with respect to $r^2$ rather than $r$, which results in a Gaussian rather than exponential decay of the solution.

\section{The $Z_N$ preserving perturbation.}
The perturbation considered in the last section was the sumplest potential that breaks the $SO(3)$ as well as the $Sdiff(2)$ symmetries but leaves an $O(2)$ subgroup of $SO(3)$ and  large subgroup $Sdiff(2)$, eq.(\ref{subgroup}) intact. Following the parallel with the 2+1 dimensional discussion, we expect the global symmetries to be broken down to $Z_N$ if our effective theory has a chance of mimicking some properties of $SU(N)$ gauge theories. In this section therefore we consider an additional perturbation, which  breaks the remaining $O(2)$ symmetry down to the $Z_N$ subgroup. 
We modify the Lagrangian to
\be L =\frac{1}{16e^2}F^2 - \frac{2}{e^2}\Lambda^2 (z - 1)^2 \Big[1- \mu (\psi^N + \psi^{\star N})\Big]  = \frac{1}{16e^2}F^2 - \frac{2}{e^2}\Lambda^2 (z - 1)^2 \Big[1- 2 \mu (1-z^2)^{N/2} \cos{N \chi}\Big] \label{tamlang} \ee
We will only consider regime where the minimum of the potential is at $z=1$. It is easy to see that this is the case as long as 
\be \mu<\frac{1}{2}\ee
For field configurations which do not depend on the longitudinal coordinate $x_3$, the energy per unit length is given by
\be\label{energy}
E=\int d^2x \frac{1}{2e^2}(\epsilon_{ij}\partial_iz\partial_j\chi)^2+\frac{2}{e^2}\Lambda^2 (z - 1)^2 \Big[1- 2 \mu (1-z^2)^{N/2} \cos{N \chi}\Big]
\ee

\subsection{Perturbative solution}
Let us first consider the perturbation to be small, $\mu \ll 1$, and  find the first order corrections to the string solution of the previous section. 
We take the following ansatz for the perturbative solution:
\be
z(r, \theta) = z(r); \chi = \theta + \chi_1(r, \theta)=\theta+f(r)\sin N\theta\ee
where $z(r)$ is given by eq.(\ref{zetsifir}). This is not the most general form of the perturbation, but which nevertheless yields a solution to first order in $\mu$, as we now show. 
%This form 

%$$ \partial_r z \partial_\theta \chi - \partial_\theta z \partial_r \chi = \partial_r z_0 + \partial_r z_0 \partial_\theta \chi_1 + \partial_r z_1$$

%to first order.
To first order in $\mu$ the equation for $\chi_1$ is

\be
\frac{1}{e^2}8N^2(z')^2f\sin N\theta=\frac{1}{e^2}N\mu\Lambda^2(z-1)^2(1-z^2)^{N/2}\sin N\theta\ee
solved by
\be\label{pertsol} f(r^2)=\frac{\mu}{N}\Big[2e^{-\Lambda r^2}(1-e^{-\Lambda r^2})\Big]^{N/2}
\ee
The second minimization equation reads
\be \frac{1}{e^2}8N\Big[2z''f+z'f'\Big]=\frac{1}{e^2}4\mu\Lambda\Big[2(z-1)(1-z^2)^{N/2}-Nz(z-1)^2(1-z^2)^{N/2-1}\Big]\ee
It is straightforward to check that this equation is indeed satisfied by the perturbative solution eq.(\ref{pertsol}) and $z(r)$ given by eq.(\ref{zetsifir}).

Calculating the longitudinal electric field corresponding to this solution we find
\be F^{03}=-4\Lambda e^{-\Lambda r^2}\Big[1+\mu\Big(2e^{-\Lambda r^2}(1-e^{-\Lambda r^2})\Big)^{N/2}\cos N\theta\Big]\ee
The electric field is concentrated within the radius $\Lambda^{1/2}$ in the transverse plane, with the $Z_N$ invariant perturbation providing a slight angular modulation, as expected.

\subsection{The General Solution}
Let us now consider the minimization equations beyond perturbation theory and beyond the simple ansatz of the previos subsection.
Minimization fo the energy functional eq.(\ref{energy}) gives the following equations:
\begin{eqnarray}
&&\frac{1}{e^2}\epsilon_{ij}\partial_j\chi\partial_iF=\frac{\partial U}{\partial z}\nonumber\\
&&\frac{1}{e^2}\epsilon_{ij}\partial_jz\partial_iF=-\frac{\partial U}{\partial \chi}
\end{eqnarray}
where 
\be F\equiv \frac{1}{2}F^{03}=\epsilon_{ij}\partial_iz\partial_j\chi\ee
and $U$ is the potential energy in eq.(\ref{energy}).

Multiplying the first equation by $\partial_k z$, the second by  $\partial_k \chi$ and subtracting, we find:

\be\label{relation} \frac{1}{2e^2}\partial_k (F^2)=\partial_k U\ee
For any finite energy configuration the electric field vanishes at infinity. Since the potential $U$ appearing in eq.(\ref{energy}) also vanishes at infinity, the integration constant needed to integrate eq.(\ref{relation}) is trivial and we find
\be F^2=2e^2 U; \ \ \ \ \ F=\sqrt{2e^2U}\ee
To solve the equation it is convenient to use the coordinates $(\tau=r^2,\theta)$:
\be\label{equation} \partial_\tau z \partial_\theta \chi - \partial_\theta z \partial_\tau \chi = \sqrt{\frac{1}{2}e^2U}\ee
This equation obviously has many solutions. The infinite degeneracy follows from a symmetry of the energy functional eq.(\ref{energy}). Consider a transformation
\be\label{sdiffr} (z(x),\chi(x))\rightarrow (z(x'),\chi(x')); \ \ \ \ \ \ \frac{\partial(x'^{1}, x'^{2})}{\partial(x^1, x^2)}=1\ee
Such transformations form the group of area preserving diffeomorphisms on a plain $SDiff(R^2)$. Note that it is a diffeomorphism transformation on the coordinate space rather than on the field space, and thus is very different from $Sdiff(2)$, which we discussed in the previous section.
This transformation is clearly a symmetry of the energy functional eq.(\ref{energy}). Thus, starting from any string solution one can generates an infinite number of degenerate solutions with the help of  $SDiff(R^2)$ transformations. Note, that since the longitudinal electric field is itself invariant under eq.(\ref{sdiffr}), all these solutions have the same electric field profile.

We will discuss here two such solutions. Let us look for solution with a prescribed and simple dependence of $\chi$ on the angle : $\chi=\theta$.
Eq.(\ref{equation}) then becomes an equation for $z$:

\be \partial_\tau z = \sqrt{\frac{1}{2}e^2U} =  \sqrt{ \Lambda^2 (z - 1)^2 \Big[1- 2 \mu (1-z^2)^{N/2} \cos{N \theta}\Big] }\ee
The dependence on $\theta$ here is parametric, and so for every value of $\theta$ it is solved by
\be\tau=\int_{-1}^{z(\tau)} dz\frac{1}{ \sqrt{ \Lambda^2 (z - 1)^2 \Big[1- 2 \mu (1-z^2)^{N/2} \cos{N \theta}\Big] }}\ee
The solution has the correct asymptotics, since as $\tau\rightarrow\infty$ the function $z$ has to approach unity for the RHS to diverge. In fact it is easy to find the large distance asymptotics of the solution. When $z$ is close to unity, we can neglect the term proportional to $\mu$ in the denominator, and for the IR asymptotics we have
\be\tau=\int_{-1}^{z(\tau)} dz\frac{1}{ \sqrt{ \Lambda^2 (z - 1)^2  }}\ee
which is solved by
\be z(\tau\rightarrow\infty)=1-2e^{-\Lambda\tau}\ee
This is the same as eq.(\ref{zetsifir}), showing that the IR asymptotics of the string solution is unaffected by the $Z_N$ perturbation.

Let us now consider a solution where $z$ only has radial dependence. In this case, we have:

\be \partial_\tau z \partial_\theta \chi = \sqrt{ \Lambda^2 (z - 1)^2 \Big[1- 2 \mu (1-z^2)^{N/2} \cos{N \chi}\Big] }\ee
This can be formally solved for $\theta$ at fixed $r$:

\be \theta = \int_0^{\chi{(r,\theta)}}  \frac{z' d \chi}{\sqrt{ \Lambda^2 (z - 1)^2 \Big[1- 2 \mu (1-z^2)^{N/2} \cos{N \chi}\Big] }}\ee

The right hand side can be expressed in terms of elliptic integrals:

\be \theta =  \frac{2}{N}\frac{z'}{\sqrt{\Lambda^2 (z-1)^2 (1-2\mu (1-z^2)^{N/2})}}F(\frac{N \chi}{2}, \frac{4\mu (1-z^2)^{N/2}}{2\mu(1-z^2)^{N/2} -1})\ee

where $F(\phi,m)$ is the incomplete elliptic integral of the first kind:

\be F(\phi,m) = \int_0^\phi (1-m\sin{\theta}^2)^{-1/2} d\theta\ee

The solution has to satisfy the boundary condition
\be\chi(\theta+2\pi) = \chi(\theta)+ 2 \pi\ee

Imposing this condition gives the equation for the radial dependence of $z$.   Using the relation $F(\frac{k \pi}{2},m) = k K (m)$, where $K(m)$ is the complete elliptic integral of the first kind, we have:

\be 2\pi =  \frac{4z'}{\sqrt{\Lambda^2 (z-1)^2 (1-2\mu (1-z^2)^{N/2})}}K(\frac{4\mu (1-z^2)^{N/2}}{2\mu(1-z^2)^{N/2} -1})\ee

One can easily check, that in the infrared for $z\rightarrow 1$ the equation reduces to 
\be z'=\Lambda (1-z)\ee
and thus has the same asymptotics as in eq.(\ref{zetsifir}).

\section{Discussion}
In this paper we tried to follow the template of 2+1 dimensions and, based on a couple of simple requirements "`guess"' a scalar theory which could be a candidate of the effective theory of 3+1 dimensional gauge theories. The theory we ended up with is not entirely satisfactory, but it does have several interesting and intriguing features. 

In the Abelian limit it has a large global symmetry group, which is spontaneously broken by lowest energy classical solutions. This symmetry is not reflected in the observables which we tentatively identified with the observables of QED.  This is similar to 2+1 dimensions, where the electromagnetic field was invariant under the magnetic $U(1)$ symmetry which acted nontrivially on the vortex field. In our 3+1 dimensional model the electromagnetic field is also invariant under the action of the (large) global symmetry group $Sdiff(2)$ which is nontrivially represented on the effective scalar fields. 

Just like in 2+1 dimensions, this global symmetry is broken by the lowest energy configurations. However, the situation here is more complicated. Whereas in 2+1 dimensions the symmetry breaking pattern is the standard one, in our 3+1 dimensional model the space of vacuum configurations is very large. It includes field configurations that have nontrivial spatial dependence, and thus breaks translational invariance in addition to the global $Sdiff(2)$ symmetry. In fact, it could well be that classical analysis fails in this model quite badly. Many of the classical vacua differ from each other only in finite region of space. Generically in such a case one expects that upon quantization these configurations become connected by tunneling transitions of finite probability. Thus it is natural to expect that the quantum portrait of moduli space is significantly different from the classical one. This is a very interesting question, but it is far beyond the scope of the present work.

Upon introduction of the symmetry breaking perturbation, the model provides a simple classical description of confinement similarly to the 2+1 dimensional case. However here also there are some peculiarity. In particular, string solutions are infinitely degenerate, as static energy for configurations which depend only on two coordinates has an additional diffeomorphism invariance. This is a different invariance than in the Abelian limit, as it involves diffeomorphism transformations in coordinate space rather than in field space. Nevertheless it results in degeneracy of the solutions, although all such solutions yield the same electric field. In the sense of electric field distributions, the solution seems to be unique. This again begs the question about the behavior of such a string in a quantum theory, since it  carries a large entropy associated with large degeneracy. 

We note that the string degeneracy is lifted if one adds the standard kinetic term for the $O(3)$ sigma model fields, $\partial^\mu\phi^a\partial_\mu\phi^a$. Addition of such a term also makes our model identical with the one proposed by Faddeev and Niemi in \cite{niemi} as an effective theory of QCD.  Interestingly, the picture we suggest is quite distinct from and complementary to that of \cite{niemi}. Whereas the authors of \cite{niemi} concentrated on closed string solutions supposedly representing glueballs, our picture is that of open strings, with the endpoints corresponding to "`constituent gluons"' like in 2+1 dimensions\cite{dualreview},\cite{baruch}. The stability of closed strings in the Faddeev-Niemi model is achieved due to nontrivial twisting of the phase of the scalar field along the string. Open strings on the other hand, do not require any twist and in principle can break into shorter strings similarly to QCD. The approximate stability of relatively long strings must be due to dynamical properties of the theory which should make the endpoints sufficiently heavy\cite{greensite}.
 
Finally we note that with the standard kinetic term our model becomes very similar to $CP^1$ model, which has been recently discussed in the literature in relation to effective models of confinement\cite{cpn}.

\section{Appendix.}

In this appendix we show that the model considered in this paper does not admit two photon solutions with arbitrary polarizations.
We are looking for two photon solutions for which the electromagnetic tensor is of the form:

\be \tilde{F}_{\mu\nu}=  \partial_{[\mu}z\partial_{\nu]}\chi=A(k_{\mu}\epsilon^1_{\nu}-k_{\nu}\epsilon^1_{\mu}) \cos{kx} + B(p_{\mu}\epsilon^2_{\nu}-p_{\nu}\epsilon^2_{\mu}) \cos{px}\ee
Foer simplicity we choose the case when the first photon has momentum $k$ in $x$-direction and polarization $a$ in $y$-direction, while the second photon has momentum $p$ in $y$-direction and polarization $b$ in $z$ direction. Note that this case is not covered by our construction of two photon states in the body of the paper.

 Now, for components of $\tilde{F}_{\mu\nu}$, we have:
\be\partial_{[0}z \partial_{1]}\chi = 0 = \partial_{[1}z \partial_{3]}\chi = 0\ee
\be \partial_{[0}z \partial_{2]}\chi = ka \cos{kx} = - \partial_{[1}z \partial_{2]}\chi\ee
\be \partial_{[0}z \partial_{3]}\chi = pb \cos{px} = -\partial_{[2}z \partial_{3]}\chi \ee

Introducing new coordinates $(x,y,z,t) \rightarrow (\bar{x}=t-x, \bar{y} = t-y, \bar{t} = t, \bar{z} = z)$, and using unbarred symbols for notational simplicity, we have:

\be \label{first}\partial_{[t}z \partial_{y]}\chi =  \partial_{[t}z \partial_{z]}\chi  =  \partial_{[x}z \partial_{z]}\chi =0\ee
\be \partial_{[t}z \partial_{x]}\chi  = \partial_{[x}z \partial_{y]}\chi= -ka \cos{kx}\ee

\be\label{last} \partial_{[y}z \partial_{z]}\chi = pb \cos{py}\ee

These equations have no solutions. Assuming $\partial_tz\ne 0$, the first two equations in eq.(\ref{first}) imply $ \partial_{y}z \partial_{z}\chi- \partial_{z}z \partial_{y}\chi = 0$, which contradicts eq.(\ref{last})the last equation. Alternatively, assuming $\partial_tz= 0$, implies vanishing of either $\partial_t\chi$, or two other partial derivatives of $z$ . It is then easy to see that both these options are in conflict with the rest of the equations. The result is that a two photon state with this polarization pattern cannot be constructed in this model. 

The model also contains solutions which do not satisfy the homogeneous Maxwell equation. As an example of such a solution consider the configuration
\be \chi=\sin p\cdot x; \ \ \ z=\sin k\cdot x\ee
It is easy to see that this configuration satisfies equations of motion, provided
\be (p\cdot k)^2-p^2k^2=0\ee
A simple example is a lightlike momentum $k^\mu$ and a spacelike momentum $p^\mu$ satisfying $p\cdot k=0$. This yields the dual field strength
\be \tilde F_{\mu\nu}\propto (k_\mu p_\nu-k_\nu p_\mu)[\cos (p+k)\cdot x+\cos (p-k)\cdot x]\ee
which is not conserved 
\be \partial^\mu\tilde F_{\mu\nu}\propto p^2 k_\nu [\sin (p+k)\cdot x+\sin(p-k)\cdot x]
\ee
In fact, both momenta $k+p$ and $k-p$ are spacelike, and thus $\tilde F_{\mu\nu}$ looks tachyonic. However, as mentioned in the Discussion, since the model classically  has many degenerate vacua with broken translational invariance, interpretaton of classical solutions as excitations is not so clear. 

\section*{Acknowledgments}
The research was supported by the DOE grant DE-FG02-92ER40716. We thank Alexei Yung and Michael Lublinsky for useful conversations.

\end{document}